\documentstyle[11pt]{article}
\newcommand{\be}{\begin{equation}}
\newcommand{\ee}{\end{equation}}

\newcommand{\bea}{\begin{eqnarray}}
\newcommand{\eea}{\end{eqnarray}}
\newcommand{\mbold}[1]{\mbox{\boldmath$#1$}}
\newcommand{\bnabla}{\mbold{\nabla}}
\title{Extended Thomas-Fermi approximation to the one-body density matrix}

\author{V.B.Soubbotin \\ {\small  Nuclear Physics
Department, Physical Research Institute},\\ {\small St.Petersburg
University, St.Petersburg, Russia} \\
{X.Vi\~nas}\\ {\small Departament d'Estructura i
Constituents de la Mat\`eria, Facultat de F\'{\i}sica,}
\\{\small Universitat de Barcelona, Diagonal 645 E-08028 Barcelona, Spain
}}
\begin{document}

\maketitle

\begin{abstract}
The one-body density matrix is
derived within the Extended Thomas-Fermi approximation. This has been
done starting from the Wigner-Kirkwood distribution function for a
non-local single-particle potential. The links between this new
approach to the density matrix with
former ones available in the literature are widely discussed. The
semiclassical Hartree-Fock energy at Extended Thomas-Fermi level is
also obtained in the case of a non-local one-body Hamiltonian.
Numerical applications are performed using the Gogny and Brink-Boeker
effective interactions. The semiclassical binding energies and root
mean square radii are compared with the fully quantal ones and
with those obtained using the Strutinsky averaged method.

\end{abstract}

\pagebreak

\section{Introduction}

The one-body density matrix (DM) $\rho (\mbold{r},\mbold{r'}) =
\sum_{\alpha} {\phi^{*}}_{\alpha}(\mbold{r}) \phi_{\alpha}
(\mbold{r'})$
or equivalently its Wigner transform the distribution function
$f(\mbold{R},\mbold{p})$ (defined
below), plays a crucial role in the Hartree-Fock (HF) calculations.
If zero-range Skyrme
forces \cite{Vauth} are used, only the diagonal part of the DM is
needed. However, full knowledge of $\rho(\mbold{r},\mbold{r'})$
(or $f(\mbold{R},\mbold{p})$) is necessary if one considers
finite-range effective nuclear forces which are derived from G-matrix
calculations in nuclear matter through the local density
approximation \cite{Neg,CaSp,Bert,Toki} or
postulated empirically with their parameters fitted to
reproduce some properties of nuclear matter and finite nuclei
\cite{BB,Gog}.

The full calculation of the density matrix (or the distribution
function) is not an easy task and requires some computational effort
\cite{Gog,MM}. Consequently, approximations which simplify the
calculation and, at the same time, show more clearly the physical
content of the DM are in order. The simplest one is to replace the
non-diagonal part of the DM by its value in nuclear matter
(Slater approach). Finite size effects are added using
the density matrix expansion (DME), either that due to
Negele and
Vautherin (NV) \cite{NV,Spru} or the modified expansion due to
Campi and Bouyssy
(CB) \cite{CB}. Very recently, the CB
approach has been applied to HF calculations of finite nuclei
\cite{Hof} using a density-dependent version of the M3Y interaction
\cite{Bert,Toki}.

On the other
hand, semiclassical methods \cite{Ring} are very useful for describing
nuclear properties of a global character such as binding energies or
nuclear densities and their moments. Concerning the nuclear ground
state properties at HF level, semiclassical approaches are based on
the
Wigner-Kirkwood (WK) $\hbar$-expansion of the distribution function which for
a set of nucleons moving in a local external potential $V(\mbold{r})$ up to
second order is given by \cite{Ring}:
\bea
f_{WK}(\mbold{R},\mbold{p}) &=& \Theta (\lambda - H_W) -
\frac{\hbar^2}{8m} \Delta V \delta' (\lambda - H_W)
\nonumber
\\
&+&
\frac {\hbar^2}{24m} \big[(\bnabla V)^2 + \frac{1}{m} (\mbold{p} .
\bnabla )^2 V \big] \delta'' (\lambda - H_W ) + {\cal{O}}(\hbar^4),
\label{eq1} \eea
where $\lambda$ is the chemical potential and $H_W$ is the Wigner transform
\cite{Ring} of the one-body Hamiltonian, which reads
\be
H_W = \frac{p^2}{2 m} + V(\mbold{R}).
\label{eq1a} \ee
 The semiclassical distribution function $f(\mbold{R},\mbold{p})$ is a
representation of the true phase-space function
in terms of distributions and
is very efficient in order to obtain semiclassical expectation values
by integrals over the whole phase-space \cite{Kri,Cen}.

The main purpose of this paper is to derive the explicit expression
of the DM in the Extended Thomas-Fermi (ETF) approximation
\cite{Brack} starting from the very recently
presented WK expansion
up to $\hbar^2$ order of the distribution function for non-local
potentials \cite{Cen}. On one hand, we want to establish a link
between the NV (and CB) expansions of the DM with this
semiclassical approach and, on the other, to apply this ETF DM to
derive the exchange HF energy when finite range forces are used.
The paper is organized as follows:
In the first section we compare the semiclassical ETF density
matrix with the former approximations of NV and CB in the case of a local
potential. In the second section
we derive the density matrix and the HF energy in the ETF approximation for a
non-local potential. We also perform restricted HF variational
calculations for some selected spherical nuclei using the Gogny
\cite{Gog} and Brink-Boeker \cite{BB} effective forces. We compare
these HF ETF results with those obtained quantally, with those obtained
with the Strutinsky average method \cite{Strut} and with those which
result from the NV and CB approaches to the DM.
In the last section we give our
conclusions and outlook. Technical details concerning the calculation of
the DM and HF energy in the ETF approach for a
non-local potential are given in the Appendix.

\section{Extended Thomas-Fermi Density Matrix}

The first step is to perform the inverse Wigner transform of (\ref{eq1})
to obtain the semiclassical WK density matrix in coordinate space.
The definition used here for the Wigner
transform of the one-body density matrix  is \cite{Ring}:
\be
f(\mbold{R},\mbold{p}) = \int d \mbold{s} e^{-i \mbold{p}
\mbold{s}/\hbar}
\rho (\mbold{R}+\frac{\mbold{s}}{2},\mbold{R}-\frac{\mbold{s}}{2}),
\label{eq2} \ee
where $\mbold{R}=(\mbold{r_1}+\mbold{r_2})/2$, $\mbold{s}=\mbold{r_1}-
\mbold{r_2}$ and $\mbold{p}$ are, respectively, the centre-of-mass,
the relative coordinates and the phase-space momentum.

After some lengthy but straightforward algebra the semiclassical
DM in terms of $\mbold{R}$ and $\mbold{s}$ at WK level is given by:

\bea
\rho(\mbold{R},\mbold{s}) &=& \frac {g k_F^3}{6 \pi^2} \frac{3
j_1(k_F s)}
{k_F s} + \frac{g}{24 \pi^2} \Delta k_F \big[ j_0 (k_F s) - k_F s j_1
(k_F s)
\big]\nonumber \\ &+& \frac{g}{48 \pi^2} \frac {(\bnabla k_F)^2}{k_F}
\big[
j_0 (k_F s) - 4 k_F s j_1 (k_F s) + k^2_F s^2 j_2 (k_F s) \big]
\nonumber \\ &-& \frac{g}{48 \pi^2} \frac{1}{k_F}\bnabla \big[ k_F
\bnabla k_F
\frac{\mbold{s}}{s} \big] \frac{\mbold{s}}{s} \big[ -3 k_F s j_1 (k_F s)
+ k^2_F s^2 j_2 (k_F s) \big] + {\cal{O}}(\hbar^4),
\label{eq3} \eea
where $k_F = \sqrt{\frac {2 m}{\hbar^2}(\lambda - V(\mbold{R}))}$ is
the local
Fermi momentum, $j_l (k_F s)$ are the spherical Bessel functions and
g stands for the degeneracy.

This expression, although written in a slightly different way, coincides with
the ones obtained previously by Dreizler and Gross \cite{Dreiz} and Jennings
\cite{Jenn}. The first term of the expansion (\ref{eq3}) corresponds to the
Slater approach, whereas the $\hbar^2$ terms are the part that take into
account quantal finite-size effects.

The WK density matrix in coordinate space depends on the angle between
$\mbold{R}$ and $\mbold{s}$, however for practical purposes and following
previous literature \cite{NV,CB,Spru} we perform the angular
average of eq.(\ref{eq3}) obtaining:
\bea
\tilde{\rho}(\mbold{R},s) &=& \tilde{\rho_0}(\mbold{R},s) + \tilde{\rho_2}
(\mbold{R},s) \nonumber \\ &=& \frac {g k_F^3}{6 \pi^2} \frac{3
j_1(k_F s)}
{k_F s} + \frac{g}{144 \pi^2} \Delta k_F \big[ 6 j_0 (k_F s) - 3 k_F s
j_1 (k_F s) - k^2_F s^2 j_2 (k_F s) \big] \nonumber \\ &+&
\frac{g}{144 \pi^2}
\frac {(\bnabla k_F)^2}{k_F} \big[ 3 j_0 (k_F s) - 9 k_F s j_1 (k_F s)
 + 2 k^2_F s^2 j_2 (k_F s) \big].
\label{eq4} \eea
The diagonal part (s=0) of eq.(\ref{eq4}) is the well-known WK
expression for the local density (with degeneracy g) \cite{Ring,Cen}:
\be
\rho(\mbold{R}) = \frac{g k_F^3}{6 \pi^2} + \frac{g}{48 \pi^ 2} \big[
\frac{(\bnabla{k_F})^ 2}{k_F} + 2 \Delta k_F \big].
\label{eq4a} \ee

To obtain the DM in the ETF approach we shall express the Fermi momentum
and its derivatives in terms of the local density and its gradients.
First, the local Fermi momentum is obtained by inverting eq.(\ref{eq4a}):
\be
k_F = k_0 - \frac{1}{24} \big[ \frac{(\bnabla{k_0})^
2}{k_0^3} + 2\frac {\Delta k_0}{k_0^ 2} \big],
\label{eq4b} \ee
where $k_0=(6 \pi^2 \rho/g)^{1/3}$. Notice that for inverting the
gradient terms in eq.(\ref{eq4a}) it is enough to replace $k_F$ by
$k_0$ to be consistent with the $\hbar$-order in the expansion
of the Fermi momentum (\ref{eq4b}). Writing the gradients of $k_0$ in
terms of the spatial derivatives of the local density
\be
k_0 (\bnabla{k_0})^2 = \frac{2 \pi^2}{3g}
\frac{(\bnabla{\rho})^2}{\rho}
\label{eq6a} \ee
\be
k_0^2 \Delta k_0 = \frac{2 \pi^2}{g} \big[ \Delta \rho - \frac{2}{3}
\frac
{(\bnabla{\rho})^2}{\rho} \big]
\label{eq6b} \ee
one finally obtains the Fermi momentum as:
\be
k_F = (\frac{6 \pi^2 \rho}{g})^{1/3} + \frac{1}{72} (\frac{6 \pi^2 \rho}{g})^
{-1/3} \big[ \frac{(\bnabla{\rho})^2}{\rho^2} - 2 \frac{\Delta \rho}{\rho}
\big],
\label{eq5} \ee
where the first term of the right-hand side is the pure Thomas-Fermi
part
and the second term, which contains derivatives of the local density,
is the $\hbar^2$ contribution.
The semiclassical density matrix for a local potential in the ETF approach
is obtained from (\ref{eq4}) by expanding consistently the Fermi
momentum $k_F$
up to $\hbar^2$-order with the help of eqs.(\ref{eq6a} - \ref{eq5}):
\bea
{\tilde{\rho}}_{ETF}(\mbold{R},s) &=& \rho \frac
{3 j_1 (k_F s)}{k_F s} + \frac {s^2}{72} \Delta \rho \big[ j_0 (k_F s) -
6 \frac {j_1 (k_F s)}{k_F s} \big] \nonumber \\ &-& \frac{s^2}{216}
\frac {(\bnabla \rho)^2}{\rho} \big[ 4 j_0 (k_F s) - 9 \frac {j_1 (k_F s)}
{k_F s} \big],
\label{eq7} \eea
where now $k_F=k_0=(6 \pi^2 \rho / g)^{1/3}$.

Let us now analyze the main properties of this semiclassical approach
as compared with the quantal case. Following refs.\cite{MM,NV}, the
quantal
DM averaged over the $\mbold{s}$ direction can be approximated by:
\be
\rho(\mbold{R},s) = \frac{1}{4 \pi} \int d \Omega e^{i \mbold{s} \hat{\mbold
{p}}/\hbar} \rho(\mbold{r},\mbold{r'})
\vert_{\mbold{r}=\mbold{r'}=\mbold{R}} =
j_0 (s \frac{\hat{p}}{\hbar}) \rho(\mbold{r},\mbold{r'})
\vert_{\mbold{r}=\mbold{r'}
=\mbold{R}},
\label{eq8} \ee
where $\hat{\mbold{p}}= -i \hbar (\bnabla_1 - \bnabla_2)/2$ is the relative
momentum operator. Expanding the Bessel function in a Taylor series
one gets:
\be
\rho(\mbold{R},s) = \sum^{\infty}_{n=0} \frac{(-1)^n s^{2n}}{(2n+1)!}
(\frac {\hat{\mbold{p}}}{\hbar})^{2n} \rho(\mbold{r},\mbold{r'})
\vert_{\mbold{r}=\mbold{r'}
=\mbold{R}} = \sum^{\infty}_{n=0} \frac{(-1)^n s^{2n}}{(2n+1)!}
M_{2n},
\label{eq9} \ee
where $M_{2n}$ are the momentum weighted integrals defined as \cite{MM}:
\be
M_{2n} = \frac{1}{(2 \pi \hbar)^3} \int d \mbold{p}
(\frac{\mbold{p}}{\hbar})^{2n}
f(\mbold{R},\mbold{p}) = (\frac {\hat{\mbold{p}}}{\hbar})^{2n}
\rho(\mbold{r},\mbold{r'})
\vert_{\mbold{r}=\mbold{r'}=\mbold{R}}.
\label{eq10} \ee
It should be pointed out that series (\ref{eq9}) as it stands is not
useful because it converges very slowly and cannot be truncated for large
s-values if the even moments are different from zero.

At this point there are two possibilities for approximating the exact DM.
One is to sum the series (\ref{eq9}) using some approach for
evaluating the
momentum weighted integrals and the other is to rearrange the series
(\ref{eq9}) in such a way that truncation is possible.

First of all, we will show that the semiclassical ETF approach to the DM eq.(
\ref{eq7}) corresponds to the whole sum of the series eq.(\ref{eq9})
if the moments
$M_{2n}$ are calculated in the ETF approximation. Starting from the
semiclassical distribution function, eq.(\ref{eq1}), the WK momentum
weighted integrals up to $\hbar^2$-order are easily derived. Expanding
consistently the Fermi momentum with the help of (\ref{eq5}) and using
eqs.(\ref{eq6a}-\ref{eq6b}) one finally obtains the ETF weighted
integrals in terms of the local density and its gradients:
\be
M^{ETF}_{2n} = \frac {3}{2n+3} \rho {k_F}^{2n} + n {k_F}^{2n-2} \big[ \frac
{8n-5}{108} \frac {(\bnabla \rho)^2}{\rho} - \frac{2n-5}{36} \Delta
\rho \big],
\label{eq11} \ee
where again $k_F=k_0=(6 \pi^2 \rho/g)^{1/3}$.

The first term of the right-hand side of (\ref{eq11}) is the
Thomas-Fermi
weighted integral while the second term is just the $\hbar^2$
correction in the ETF approach. For $n=0$ and $n=2$ one obtains:
\be
M^{ETF}_0 = \rho
\label{eq12} \ee
\be
M^{ETF}_2 = \frac{3}{5} \rho {k_F}^2 + \frac{1}{36} \frac{(\bnabla \rho)^2}
{\rho} + \frac{1}{12} \Delta \rho,
\label{eq13} \ee
which are just the semiclassical counterparts (at ETF-$\hbar^2$ level)
of the
zeroth and second-order quantal momentum weighted integrals: $M_0 = \rho$ and
$M_2 = \tau- \Delta \rho/4$, where $\tau$ is the kinetic energy
density. Notice
that in the ETF-$\hbar^2$ approach only second-order gradients of the
local
density appear in $M^{ETF}_{2n}$ for any value of $n$. However, higher order
derivatives will appear in the moments if the ETF expansion is pushed
to higher powers in $\hbar$.

Taking into account the Taylor expansion of the Bessel functions in eq.
(\ref{eq7}) and after some algebra one finds:
\bea
\tilde{\rho}_{ETF}(\mbold{R},s) &=& \rho \frac {3 j_1 (k_F s)}{k_F s}
- \sum^{\infty}_{n=0} \frac{(-1)^n s^{2n+2}}{(2n+3)!} (n+1) {k_F}^{2n}
\big[
\frac {8n+3}{108} \frac {(\bnabla \rho)^2}{\rho}
- \frac{2n-3}{36} \Delta \rho \big]
\nonumber \\
&=& 3 \rho \sum ^{\infty}_{n=0} \frac{(-1)^n (k_F s)^{2n}}
{(2n+3)(2n+1)!} + \sum^{\infty}_{n=1} \frac {(-1)^n s^{2n}}{(2n+1)!}
\big[ M^{ETF}_{2n} - \frac {3}{2n+3} \rho {k_F}^{2n} \big]
\nonumber \\
&=& \sum^{\infty}_{n=0}
\frac {(-1)^n s^{2n}}{(2n+1)!} M^{ETF}_{2n}.
\label{eq14} \eea
From this result it is clear that in the ETF approximation to the DM
all the momentum
weighted integrals appearing in eq.(\ref{eq9}), evaluated within the
same semiclassical approach, are consistently summed.

Another possibility for approximating the quantal DM is to rearrange
the terms in
eq.(\ref{eq9}) in such a way that the leading term is the Slater term.
This is, actually, the way in which the NV and CB approaches to
the DM are done. For the sake of completeness we shall once again
briefly derive the NV and CB approaches to the DM following the
method outlined in ref. \cite{MM}.

The starting point is the the identity
\be
j_0 (ab) = \sum^{\infty}_{n=0} (-1)^n (4n+3) \frac {j_{2n+1} (ak)}{ak}
\frac {P_{2n+1} (b/k)}{b/k}
\label{eq15} \ee
valid for any $k$ such that $-1 \le b/k \le 1$ and where $P_{2n+1}(x)$ are the
Legendre polynomials $P_{2n+1}(x) = \sum^n_{m=0} a^n_m x^{2(n-m)+1}$ with
\be
a^n_m = \frac{(-1)^m}{2^{2n+1}} \frac {(4n-2m+2)!}{m! (2n-m+1)! (2n-2m+1)!}.
\label{eq16} \ee
Using eq.(\ref{eq15}) in eq.(\ref{eq8}), the angular averaged DM eq.(\ref{eq9}
) is also written as:
\bea
\rho(\mbold{R},s) &=& \sum^{\infty}_{n=0} (-1)^n (4n+3) \frac {j_{2n+1} (ks)}
{ks} \frac {P_{2n+1} (\hat{p}/\hbar k)}{\hat{p}/\hbar k}
\rho(\mbold{r},\mbold{r'})
\vert_{\mbold{r}=\mbold{r'}=\mbold{R}} \nonumber \\ &=&
\sum^{\infty}_{n=0}
\frac {(-1)^n \hat{j}_{2n+1} (ks) s^{2n}}{(2n+1)!} \sum^n_{m=0} \frac
{a^n_m}
{a^n_0} M_{2(n-m)} k^{2m},
\label{eq17} \eea
where $\hat{j}_{2n+1}(ks)=(4n+3)!! j_{2n+1}(ks)/(ks)^{2n+1}$ are the spherical
Bessel functions normalized to unity at $s=0$.

Of course, the semiclassical $\tilde{\rho}_{ETF}(\mbold{R},s)$
(\ref{eq14}) also fulfills eq.(\ref{eq17}). Starting from this
equation if the moments are obtained in the ETF approximation
(\ref{eq11}), the Bessel functions
$\hat{j}_{2n+1}(ks)$ are expanded in a Taylor series and the even
powers of $s$ are properly sorted out, one recovers (\ref{eq14}).

The NV approach consists of keeping the two first terms of the
expansion
(\ref{eq17}) and taking $k=k_F$. In this case the DM reads:
\be
\tilde{\rho}_{NV}(\mbold{R},s) = \rho \frac {3 j_1 (k_F s)}{k_F s}
- \frac{35}{2} \frac{j_3 (k_F s)}{k_F^3 s} \big[ \frac {3}{5} k_F^2 M_0
- M_2 \big]
\label{eq18} \ee
The CB approximation keeps only the first term of (\ref{eq17}) but with k
fixed in such a way that the second term of (\ref{eq17}) identically vanishes:
\be
\tilde{\rho}_{CB}(\mbold{R},s) = \rho \frac {3 j_1 (\tilde{k}_F s)}
{\tilde{k}_F s},
\label{eq19} \ee
where $\tilde{k}_F = \sqrt{5 M_2/3\rho}$.

The NV and CB approaches are truncations of the true expansion of the quantal
DM eq.(\ref{eq9}). As is discussed in \cite{MM}, in these approximations
only the $M_0$ and $M_2$ momentum weighted integrals correspond to
those
obtained with the exact DM, whereas any higher momentum weighted
integral in
these approaches $M_{2 \lambda}$ ($\lambda > 1$) has little to do with its
exact quantal value.

Let us now discuss the results obtained using the different approaches
to the DM analyzed previously. To do this we consider a $^{40}Ca$
nucleus described using harmonic
oscillator (HO) wavefunctions with an oscillator parameter $\alpha =\sqrt{m
\omega/\hbar}$ = 0.516 fm$^{-1}$.
Figure 1 displays the ratio of the off-diagonal to diagonal DM $\rho(\mbold{R},s)
/\rho(\mbold{R})$ as a function of the interparticle distance $s$ for selected
values of the distance from the centre-of-mass $\mbold{R}$. The
different
curves shown in this figure correspond to the quantal DM (black dots)
, the NV (dashed-dotted line) and CB (dashed line) approximations and
the semiclassical ETF approach calculated using the quantal local density
(solid line). As general trends the approximations to the DM
considered here reproduce reasonably well the quantal values in the range of
$R=1-3$ fm, they show some deficiencies at $R=0$ fm and clearly start to
separate from the quantal values for $R>4$ fm. However, in the whole range of
$R$-values analyzed, the quantal approaches NV and CB reproduce the quantal
DM better than the semiclassical ETF calculation.

At this point two comments are
in order. First of all, it should be pointed out that all the approaches to
the DM considered in this paper are, in fact, distributions
(see eq.(\ref{eq1})
for the ETF DM and ref. \cite{MM} for the discussion of the NV and CB
cases) and consequently,
the only meaningful comparison should be done through the moments in $k$ and
$R$ spaces.

On the other hand, the semiclassical approaches to the DM are obtained
by switching off shell effects.
Consequently, the Slater and ETF approximations to the DM should be compared
with the smoothed DM obtained using the Strutinsky averaged occupation numbers
\cite{Strut} rather than with the quantal DM. To do this, one starts
from the
smooth distribution function, which for closed HO shells reads
\cite{Prak}:
\be
\tilde{f}_{St}(\mbold{R},\mbold{p}) = 8 \sum^{\infty}_K (-1)^K \tilde{n}_K e^{-
\varepsilon} L^2_K(2 \varepsilon),
\label{eq20} \ee
where $\varepsilon = p^2/m +m \omega^2 R^2$, $L^{\alpha}_K$ are the
generalized
Laguerre polynomials and $\tilde{n}_K$ the Strutinsky occupation numbers
\cite{Strut}. Performing the inverse Wigner transform of (\ref{eq20})
and
averaging over the angles, one obtains the Strutinsky DM in coordinate space
to which the semiclassical approximations (Slater and ETF) should be compared.
\bea
\tilde{\rho}_{St}(\mbold{R},s) &=& \
\int \frac{d \mbold{p}}{(2 \pi \hbar)^3}
\tilde{f}_{St}
(\mbold{R},\mbold{p}) e^{i \mbold{p} \mbold{s}/\hbar3} \nonumber \\ &=&
\frac{\alpha^3}{\pi^{3/2}} \sum^{\infty}_{K=0} \tilde{n}_K e^{- \alpha^2 (R^2
+\frac{s^2}{4})} \sum^K_{K_1=0}(-1)^{K_1} L^{1/2}_{K_1} (2 \alpha^2 R^2 )
L^{1/2}_{K-K_1}(\frac{\alpha^2 s^2}{2}).
\label{eq21} \eea
The NV and CB expansions of the DM can also be considered within the
semiclassical
framework if the $M_2$ moment is calculated in the ETF approach.
From (\ref{eq14}), it is clear that in this case NV and CB become
truncations of the ETF density matrix. It is interesting to look at
the quality of these approximations because they
have been used
for calculating the exchange part of the nucleus-nucleus potential (see \cite
{Ism} and references quoted therein).

The Strutinsky DM (\ref{eq21}) for $^{40}Ca$ is obtained with a HO
parameter $\alpha$=0.516 $fm^ {-1}$ and with a smoothing parameter \cite
{Ring,Strut} $\gamma$=1.25 $\hbar \omega$.
Figure 2 collects the
semiclassical results where the ratio $\tilde{\rho}(\mbold{R},s)/\rho(\mbold
{R})$ is displayed for the Strutinsky smoothed DM (\ref{eq21}) (black dots)
and for the ETF DM (solid line) as well as for the
semiclassical NV
(dashed line) and CB (dashed-dotted line) truncations of the ETF DM.
Notice that in order to completely remove the shell effects, the ETF,
NV and CB
density matrices have to be obtained using the Strutinsky
local density \cite{Bra76}. From this Figure it can be seen that the ETF
ratio reproduces reasonably well the Strutinsky result. The ETF quotient
is better than the NV result in the whole range of $R$ and $s$
distances analyzed and better than the CB for small values of $R$. The
difference between ETF and Strutinsky ratios indicates that the $\hbar$-
expansion in ETF does not fully converged. Consequently, it would be
necessary to add $\hbar^4$ contributions to the ETF DM to obtain ETF
expectation values in good agreement with the corresponding Strutinsky
results.

\section{Restricted variational energy calculations}

The second part of this paper is devoted to discussion of the ability
of the ETF
approach to the DM for describing the HF binding energy of finite nuclei when
effective finite-range nuclear interactions are considered.

First, we derive the ETF approximation to the DM starting from the
non-local HF potential:
\be
V^{HF}(\mbold{r},\mbold{r'}) = V^H(\mbold{r},\mbold{r'}) \delta (\mbold{r}-
\mbold{r'}) - V^F(\mbold{r},\mbold{r'}),
\label{eq22} \ee
where $V^H$ and $V^F$ are the direct and exchange parts of the HF potential.
In Wigner representation eq.(\ref{eq22}) becomes:
\be
V (\mbold{R},\mbold{k})= V_{dir}(\mbold{R}) -
V_{ex}(\mbold{R},\mbold{k}),
\label{eq22a} \ee
where
\be
V_{dir}(\mbold{R}) = \int d \mbold{R'} v(\mbold{R},\mbold{R'}) \rho (\mbold{R'})
\label{eq22b} \ee
and
\be
V_{ex}(\mbold{R},\mbold{k}) = g \int \frac{d \mbold{k'}}{(2 \pi)^3}
w(\mbold{k},\mbold{k'}) f(\mbold{R},\mbold{k'}).
\label{eq22c} \ee
In these equations $v(\mbold{R},\mbold{R'})$ is the two-body effective
interaction, $w(\mbold{k},\mbold{k'})$ is its Fourier transform and g stands for
the degeneracy (for the sake of simplicity we consider a simple Wigner force
in eqs.(\ref{eq22b}) and (\ref{eq22c})).
Consequently, the Wigner transform of the one-body Hamiltonian will
be;
\be
H_W (\mbold{R},\mbold{k}) = \frac{\hbar^2 k^2}{2 m} +  V(\mbold{R},\mbold
{k}).
\label{eq23} \ee
If the HF potential is spherically symmetric in $\mbold{k}$, i.e.
$V(\mbold{R},k)$, the WK distribution function required for
semiclassical calculations is \cite{Cen};
\bea
\tilde{f}_{WK}(\mbold{R},k) &=&
\frac{1}{4 \pi} \int f_{WK}(\mbold{R},\mbold{k}) d \Omega_k  \nonumber \\
&=& \Theta (\lambda - H_W)
- \frac{\hbar^2}{8 m} \delta'(\lambda - H_W) F_1(\mbold{R},k)
+ \frac{\hbar^2}{24 m} \delta''(\lambda - H_W) F_2(\mbold{R},k),
\label{eq24} \eea
where the functions $F_1(\mbold{R},k)$ and $F_2(\mbold{R},k)$ are given by:
\be
F_1(\mbold{R},k) = \frac{\hbar^2}{3 m} \big[ \frac{m}{\hbar^2} \Delta V
(3f + k f_k) -k^2 (\bnabla{f})^2 \big]
\label{eq24a} \ee
\be
F_2(\mbold{R},k) = \frac{\hbar^2}{3 m} \big[ \frac{m}{\hbar^2} (\bnabla{V})^2
(3f + k f_k) + k^2 f^2 \Delta V -2 k^2 f \bnabla{V} \bnabla{f} \big].
\label{eq24b} \ee
In eqs.(\ref{eq24a}- \ref{eq24b}), $f$ is the inverse of the position and
momentum dependent effective mass:
\be
f(\mbold{R},k) = \frac {m}{m^{*}(\mbold{R},k)}
= 1 + \frac {m}{\hbar^2 k} V_{ex,k} (\mbold{R},k),
\label{eq25} \ee
where the suscript $k$ indicates a partial derivative with repect to
$k$.

Due to the fact that the effective-mass corrections are included in
the
$\hbar^2$ part
of the distribution function (\ref{eq24}), they are calculated using
the $\hbar^0$ order exchange potential in eq.(\ref{eq25}) to be
consistent with the $\hbar$-order in the expansion of the WK
distribution function.

Following the steps indicated in the Appendix, the ETF density matrix for each
kind of nucleon in the case of a non-local potential
can be written as:
\bea
\tilde{\rho}(\mbold{R},s) & = & \rho \frac {3 j_1 (k_F s)}{k_F s}
+ \frac {s^2}{216} \bigg\{ \big[ (9 - 2 k_F \frac {f_k}{f} - 2 k_F^2
\frac {f_{kk}}{f} + k_F^2 \frac {f_k^2}{f^2}) \frac {j_1(k_F s)}{k_F
s}
- 4 j_0(k_F s) \big] \frac{(\bnabla{\rho})^2}{\rho}
\nonumber \\ [2mm]  && \mbox{}
- \big[ (18 + 6 k_F \frac{f_k}{f}) \frac{j_1(k_F s)}{k_F s} -
3 j_0(k_F s) \big] \Delta \rho
\nonumber \\ [2mm] && \mbox{}
- \big[ 18 \rho \frac{\Delta f}{f} + (18 - 6 k_F \frac{f_k}{f})
\frac{\bnabla{\rho}.\bnabla{f}}{f} + 12 k_F
\frac{\bnabla{\rho}.\bnabla{f_k}}{f} - 9 \rho
\frac{(\bnabla{f})^2}{f^2} \big] \frac{j_1(k_F s)}{k_F s} \bigg\},
\label{eq27} \eea
where now $k_F=(3 \pi^2 \rho)^{1/3}$ and the inverse effective mass
$f$ (\ref{eq25}) and its derivatives are computed at $k=k_F$.
We use here g=2 because in this way the ETF DM eq.(\ref{eq27}) can be
directly applied to non-symmetric nuclei.

If all the spatial and momentum derivatives of the inverse effective
mass are dropped in eq.(\ref{eq27}) one recovers the ETF DM for the
local case eq.(\ref{eq7}). If only the momentum derivatives of $f$
are eliminated, one obtains the ETF DM corresponding to the case of a
position-dependent effective mass. This latter case is just the
situation that appears when one uses zero-range forces such as the
Skyrme interactions.

The next step is to obtain the ETF approach to the HF energy,
which for an uncharged and spin-saturated nucleus can be written as:
\be
E_{HF} = \sum_q \int d \mbold{R} \bigg[ \frac{\hbar^2 \tau(\mbold{R})}{2 m} +
\frac{1}{2} \rho(\mbold{R}) V_{dir} (\mbold{R})  - \frac{1}{2}  \int d
\mbold{s} V_{ex}(\mbold{R},s) \rho (\mbold{R} + \frac {\mbold{s}}{2},
\mbold{R} - \frac {\mbold{s}}{2}) \bigg]_q,
\label{eq28} \ee
where the subindex q stands for each kind of nucleon.

The HF energy in the ETF approximation is obtained
as explained in the Appendix and reads:
\bea
\tilde{E}_{ETF} = \sum_q \int d \mbold{R} \bigg[ \frac{\hbar^2}{2 m}
\tau_{ETF}(\mbold{R}) + \frac{1}{2}
\rho(\mbold{R})V_{dir}(\mbold{R})
- \varepsilon_{ex}^{ETF}(\mbold{R}) \bigg]_q
\label{eq29} \eea
In this equation $\tau_{ETF}(\mbold{R})$ is the kinetic energy up to
$\hbar^2$ order for each kind of nucleon
in the ETF approximation and reads
\bea
\tau_{ETF}(\mbold{R}) &=& \tau_{ETF,0}(\mbold{R}) + \tau_{ETF,2}
(\mbold{R}) \nonumber \\ &=&
\frac{3}{5} k_F^2 \rho + \frac {1}{36} \frac {(\bnabla{\rho})^2}{\rho}
\big[ 1 + \frac{2}{3} k_F \frac{f_k}{f} + \frac{2}{3} k_F^2
\frac{f_{kk}}{f} - \frac{1}{3} k_F^2 \frac {f_k^2}{f^2} \big] +
\frac{1}{12} \Delta \rho \big[ 4+ \frac{2}{3} k_F \frac{f_k}{f} \big]
\nonumber \\ &+& \frac{1}{6} \rho \frac{\Delta f}{f} + \frac{1}{6}
\frac{\bnabla{\rho}.\bnabla{f}}{f} \big[ 1 - \frac{1}{3} k_F
\frac{f_k}{f} \big] + \frac{1}{9}
\frac{\bnabla{\rho}.\bnabla{f_k}}{f}
- \frac{1}{12} \rho \frac{(\bnabla{f})^2}{f^2}
\label{eq30} \eea
and $\varepsilon_{ex}^{ETF}(\mbold{R})$ is the exchange
energy density for each kind of nucleon in the same approximation given
by
\bea
\varepsilon_{ex}^{ETF}(\mbold{R}) &=&
\varepsilon_{ex,0}^{ETF}(\mbold{R}) +
\varepsilon_{ex,2}^{ETF}(\mbold{R})
= \frac{1}{2} \rho (\mbold{R}) \int d \mbold{s}
v(s) \frac{9 j_1^2 ( k_F s)}{k_F^2 s^2} \nonumber \\
&+& \frac {\hbar^2}{2 m} \big[ (f - 1)
(\tau_{ETF} - \frac{3}{5} k_F^2 \rho - \frac{1}{4} \Delta \rho)
+ k_F f_k (\frac{1}{27} \frac{(\bnabla{\rho})^2}{\rho} - \frac{1}{36}
\Delta \rho) \big],
\label{eq33} \eea
where $v(s)$ is the central nucleon-nucleon interaction,
 $k_F=(3 \pi^2 \rho)^{1/3}$ and $f$ and its k derivatives
are calculated at $k=k_F$.
For Gaussian type forces such as the Gogny or Brink-Boeker interactions
used in this
paper, the explicit form of the lowest order exchange energy
$\varepsilon_{ex,0}^{ETF}(\mbold{R})$ can be found in ref.
\cite{Soub}.

In the special case of a local potential the $\hbar^2$ part of
the kinetic energy density eq.(\ref{eq30}) reduces to
the well-known result
\be
\tau_{ETF,2} = \frac{1}{36} \frac{(\bnabla \rho)^2}{\rho} +
\frac{1}{3}\Delta \rho,
\label{eq33a} \ee
if the effective mass is not momentum dependent, one recovers the
result of \cite{Brack}
\be
\tau_{ETF,2} = \frac{1}{36} \frac{(\bnabla \rho)^2}{\rho} +
\frac{1}{3}\Delta \rho
+ \frac{1}{6} \frac{\rho \Delta f}{f}
+ \frac{1}{6} \frac{\bnabla \rho \bnabla f}{f}
- \frac{1}{12} \frac{\rho (\bnabla f)^2}{f^2}.
\label{eq33b} \ee
For the particular case of the Coulomb potential, the direct calculation of the
exchange energy density up to $\hbar^2$ order in the ETF approach
(local case) leads to
\be
\varepsilon_{Coul,ex}^{ETF} = - \frac{3}{4} (\frac{3}{\pi})^{1/3}
\rho^{4/3}
- \frac{7}{432 \pi (3 \pi^2)^{1/3}}
\frac{(\bnabla \rho)^2}{\rho^{4/3}}
\label{eq34} \ee
where $\rho$ is the proton density. Eq.(\ref{eq34}) agrees with
the result reported previously in \cite{Dreiz}.

As a first test of our ETF approach, let us compare the exchange
Coulomb energy obtained using eq.(\ref{eq34}) with the quantal
result as well as the same energy derived through the NV, CB and
Slater approximations to the DM. To this end and following ref.
\cite{CB}, we use HO wavefunctions with fixed parameters
$\alpha=0.752 fm$ for $^{4}He$, 0.546 $fm$ for $^{16}O$ and 0.481
$fm$ for $^{40}Ca$. Table 1 shows the quantal (QM label), NV, CB,
Slater (SL label) and ETF results for the exchange Coulomb energy.
From this Table it can be seen that the ETF results almost reproduce
the quantal values and improve those obtained using the NV, CB and
Slater approximations.

\subsection{Comparison with quantal results}

In order to check the quality of our approach in the calculation of HF
energies, we present here a restricted variational calculation for
uncharged $^4He$, $^{16}O$ and $^{40}Ca$ nuclei using the Brink-Boeker
and Gogny forces. To this end we use HO wavefunctions and minimize
with respect to the HO parameter $\alpha=\sqrt{m \omega/\hbar}$.

At this point it should be noted that the semiclassical energy
(\ref{eq29}) (as well as that calculated in the simplest Slater
approach) is free of shell effects if it is obtained using a smooth density
\cite{Bra76}. Consequently, the semiclassical results
should be compared rather with a HF calculation based on the smoothed
Strutinsky density matrix (\ref{eq21}) than with the purely quantal HF
calculations. However, a direct comparison with quantal results in
the mean field approach is possible if shell effects are added to
the semiclassical results (at Slater or ETF levels) according to the
Strutinsky energy theorem. One possible way of incorporating shell-effects is
based on the Kohn-Sham \cite{KS} approximation widely used in
atomic physics \cite{Dreiz} and discussed for the nuclear case in \cite{Bra97}.
Basically the KS approach consists of solving the quantal HF equation
using a
local effective potential obtained as a functional derivative of the
density-dependent exchange-correlation energy. In our restricted
variational
calculation the Kohn-Sham scheme is applied by minimizing the sum of
the quantal
kinetic plus direct energies with the semiclassical (Slater or ETF)
exchange energy eqs.(\ref{eq33}).

The exact quantal energies (corrected from
the centre-of-mass motion) are obtained from eq.(\ref{eq28}) with the
DM
evaluated analytically \cite{Thier} (however,see below for the special
case of closed HO shells)
together with its corresponding root mean square radius (RMSR)
$<r^2>^{1/2}$. These quantities are
displayed in Table 2 with the label QM for the Brink-Boeker and
Gogny force. Table 2 also shows the same results obtained
using the NV and CB truncations of the quantal DM (labelled NV and
CB respectively). The binding energies and RMSR obtained within the KS
scheme starting from the semiclassical Slater or ETF exchange
energies eq.(\ref{eq33}) are also collected in
Table 2 with the SL(KS) and ETF(KS) labels.
 The differences between the KS (Slater and ETF) results and the
purely quantal ones show the quality of the semiclassical approach
to the exchange energy. From this comparison one can see that the
Slater approach is very poor in the case of the Brink-Boeker force,
underbinding all of the considered nuclei and giving RMSR larger than
the quantal values. However, the result is more satisfactory for the
Gogny force. This
difference is due to the fact that the non-local effects are larger
in the Brink-Boeker force than in the Gogny case. The non-local
effects are better accounted for in the ETF(KS) approximation for
which agreement with the quantal HF results is good for both
effective forces considered in this paper. The ETF results
in the KS scheme are similar to those obtained using the NV
and CB approximations to the quantal DM for both Brink-Boeker and
Gogny interactions.

As has been pointed out in Section 2, the NV and CB truncations
of the quantal DM become truncations of the ETF DM if the $M_2$
momentum weighted integrals are also computed with the same ETF
approach (\ref{eq13}). To check the quality of these approximations to
the ETF DM, we have again computed the binding
energies and RMSR using the NV and CB truncations of the ETF DM
within the KS scheme. The
corresponding results are also collected in Table 2 with the NV(KS)
and CB(KS) labels. From the analysis of the KS results it can be
seen that the agreement of NV(KS) and CB(KS) with ETF(KS) is
similar to the one found comparing the NV and CB results with the QM
values. On the other hand, this
agreement improves when the non-locality of the effective force is
smaller. From Table 2 it can also be seen that the ETF(KS) energies
and RMSR agree better with the corresponding quantal results
than the NV(KS) and CB(KS) values.

Some time ago another different approximation was presented in the
literature \cite{Mey}. In this approach rather than starting from
first
principles, a phenomenological density matrix is proposed in which
the parameters were determined by imposing the correct local
semiclassical kinetic energy density and the projector character of
the DM in an integrated form. Using this parametrized DM and the Gogny
interaction,
the binding energy of $^{16}O$ and $^{40}Ca$ (including the Coulomb
energy) are
$-128.3 MeV$ and $-337.9 MeV$ respectively. These results can be
compared with the ones obtained in our ETF(KS) approach $-122.6 MeV$
and
$-335.2 MeV$ and with the fully quantal results $-126.0 MeV$ and
$-338.4 MeV$. The results obtained using the ETF(KS) approximation are
slightly worse than those obtained with the phenomenological DM
discussed previously, but clearly improve the CB results reported in
\cite{Mey}, which are $-118.2 MeV$ and $-326.4 MeV$ for $^{16}O$ and
$^{40}Ca$ respectively.

\subsection{Comparison with Strutinsky results}

Let us now discuss the quality of the Slater and ETF energies within
the
semiclassical framework. In this case they have to be compared with the
ones obtained using the Strutinsky averaged method \cite{Strut}. The starting
point for a Strutinsky calculation of the energy using trial HO wavefunctions
is the smooth density matrix (\ref{eq21}) from which the particle and
kinetic
energy can also be derived. For an effective nuclear interaction with
two Gaussian type form
factors (as in the case of the forces studied in this paper) and HO
closed
shells, the direct and exchange energies can be obtained analytically:
\bea
E_{dir} = \sum_{i=1}^2 \frac{16 X_{d,i}}{\sqrt{\pi}}
\bigg(\frac{\alpha^2}{\gamma_i}\bigg)^{3/2} \sum_{K=0}^{K_{max}}
\sum_{M=0}^{K_{max}} \tilde{n}_K \tilde{n}_M \sum_{K_1=0}^K
\sum_{M_1=0}^{M}
L_{K-K_1}^{1/2}(0) L_{M-M_1}^{1/2}(0) \nonumber \\
\frac{\Gamma(K_1 + M_1 + \frac{3}{2})}{K_1 ! M_ 1 !}
\bigg(\frac{\alpha^2}
{\gamma_i} \bigg)^{K_1 + M_1} F \bigg[ - M_1, -K_1, -M_1 -K_1 -
\frac{1}{2},
1 -\frac{\gamma_i^2}{\alpha^4}\big(1 - \frac{\alpha^2}{\gamma_i}
\big)^2 \bigg]
\label{eq35} \eea
\bea
E_{exc} = \sum_{i=1}^2 \frac{32 X_{e,i}}{\pi}
\bigg(\frac{\alpha^2}{\gamma_i}\bigg)^{3/2} \sum_{K=0}^{K_{max}}
\sum_{K_1=0}^{K_{max} - K} \sum_{M_1=0}^{K_{max} - K} \tilde{n}_{K+K_1}
\tilde{n}_{K+M_1} \frac{\Gamma(K + \frac{3}{2})}{K !} \nonumber \\
\frac{\Gamma(K_1 + M_1 + \frac{3}{2})}{K_1 ! M_ 1 !}
\bigg(1 - \frac{\alpha^2}
{\gamma_i} \bigg)^{K_1 + M_1} F \bigg[ - M_1, -K_1, -M_1 -K_1 -
\frac{1}{2},
1 -\frac{\alpha^4}{\gamma_i^2}\big(1 - \frac{\alpha^2}{\gamma_i}
\big)^{-2} \bigg],
\label{eq36} \eea
where $\gamma_i = 2/\mu_i^2 + \alpha^2$ and $F$ are the Gauss
hypergeometric functions. In eqs.(\ref{eq35}-\ref{eq36})
$X_{d,i}= w_i + b_i/2 -h_i/2 - m_i/4$ and $X_{e,i}= w_i/4 + b_i/2 -
h_i/2 - m_i$ are the usual combination of the direct and exchange
parameters of the central effective interaction and $\mu_i$ is the
range of each Gaussian form factor.

The Strutinsky occupation numbers that come into in the energy
calculation are
obtained from a HO spectrum. In this way the smooth energy becomes a function
of the HO length $\alpha$. The Strutinsky energy is obtained
by minimizing with respect to $\alpha$ to simulate the
self-consistency
\cite{Bra75} with the additional constraint that the minimization
procedure is performed in the plateau region \cite{Ring,Strut}.
With a smoothing parameter $\gamma$=1.25 $\hbar
\omega$, the HO parameters that mininimize the Strutinsky energies
of the $^4He$, $^{16}O$ and $^{40}Ca$ nuclei are $\alpha$=0.647, 0.550
and 0.509 $fm^{-1}$ respectively using the Brink-Boeker force and
$\alpha$=0.643, 0.567 and 0.516 $fm^{-1}$ in the case of the Gogny
interaction.
The binding energies and RMSR
obtained in this way for uncharged $^4He$, $^{16}O$ and $^{40}Ca$ nuclei
are collected in Table 3 with the label ST. The semiclassical binding
energies at Slater and ETF levels are computed starting from the
Strutinsky particle density obtained previously
in order to drop the shell effects completely \cite{Bra76}. These
results are shown in Table 3 labelled SL and ETF(a)
respectively.

The Strutinsky value represents the energy which varies smoothly
with the number of nucleons A. For each nucleus the difference between its
quantal value QM reported in Table 2 and the corresponding ST result
given in
Table 3 is the so-called shell energy. This is a subtle quantity that
is not reproduced by SL or ETF approaches up to $\hbar^2$ order.
As has been pointed out
in previous literature, if the ETF kinetic energy density functional only
contains the $\hbar^0$ and $\hbar^2$ contributions, its integral is not able
to reproduce the Strutinsky kinetic energy at least in the case of a set of
nucleons moving in a HO or Woods-Saxon type external potential.
\cite{Bra76}. However, if the $\hbar^4$ contributions are included in the
functional, the ETF kinetic energies are in much better agreement with the
Strutinsky values \cite{Bra76,Cen90}. In our non-local calculations the
differences found between Strutinsky and ETF (up to $\hbar^2$
order) total energies are roughly similar
to those found for the kinetic energy in the case of an external HO potential
\cite{Cen90}. This fact suggests including approximately the
$\hbar^4$ corrections to ETF by adding to $\tau_{ETF}$ which enters
equations (\ref{eq30}) and (\ref{eq33}) the $\hbar^4$ contribution
in the local potential case:
\be
\tau_4 = \frac{1}{6480} (3 \pi^2)^{-2/3} \rho^{1/3} \big[ 24 (\frac{\Delta \rho}
\rho)^2 - 27 \frac{\Delta \rho}{\rho} (\frac{\bnabla{\rho}}{\rho})^2
+ 8 (\frac{\bnabla{\rho}}{\rho})^4 \big].
\label{eq37} \ee
From this approximated calculation we find that almost all the correction comes
from the kinetic energy term, whereas the exchange part gives only a minor
contribution. The total energy when this $\hbar^4$ correction is
included perturbatively
is shown in Table 3 labelled ETF($\hbar^4$).
It can be seen that the Strutinsky binding energies are very well
reproduced by this approximate ETF($\hbar^4$) calculation showing
again the importance of including $\hbar^4$ corrections in ETF
in order to obtain a better description of the shell energies
\cite{Bra76,Cen90}.

For finite-range forces the non-local effects contribute to
the DM (\ref{eq27}) through the gradients and the derivative
with respect to $k$ of the inverse effective
mass calculated at $k=k_F=(6 \pi^2 \rho/g)^{1/3}$. To
investigate the influence of these
non-local corrections to the HF energy, we have again obtained this
energy
using the DM corresponding to a local potential (\ref{eq7}).  In this
case,
the kinetic energy reduces to that corresponding to the local case and
the $\hbar^2$ exchange energy is calculated using (\ref{eq33}) but with
$\tau$ corresponding to the local case. The HF energies calculated in this
way are also displayed in Table 2 with the label ETF(b). From these
results
it can be seen that, in fact, the $\hbar^2$ effective mass corrections
to the DM (\ref{eq27}) are almost negligible for the Gogny force but they
become more important for the Brink-Boeker interaction where the
non-local effects are larger.

\section{Summary and Outlook}

In this paper we have derived, for the first time to our knowledge,
the
Extended Thomas-Fermi approximation to the one-body density matrix up
to $\hbar^2$-order for a non-local single particle Hamiltonian. The
$\hbar^2$ contribution can be written in terms of spherical Bessel
functions combined with second-order gradients of the local density
and the inverse of the effective mass as well as momentum derivatives
of the latter computed at the Fermi momentum. This density matrix
includes, as particular cases, results reported previously in the
literature \cite{Dreiz,Jenn} for the local case.

We have compared this new semiclassical approximation with former
approaches, namely the Negele-Vautherin \cite{NV} and Campi-Bouyssy
\cite{CB} ones. It is found that as in the case of the quantal density
matrix, the Extended Thomas-Fermi
approximation sums all the momentum weighted integrals \cite{MM},
but with
their quantal values replaced by their semiclassical counterparts.
In this respect the Extended Thomas-Fermi approach differs from the
Negele-Vautherin and Campi-Bouyssy approximations, which are
truncations
of the quantal density matrix. It should also be pointed out that
if within the Negele-Vautherin and Campi-Bouyssy approaches the
quantal momentum
weighted integrals are replaced by their semiclassical counterparts,
they become truncations of the Extended Thomas-Fermi density matrix.

We have applied this semiclassical approach for deriving the
Hartree-Fock energy of a nucleus in the case of effective finite-range
interactions. In this case the $\hbar^2$-order Extended Thomas-Fermi
kinetic and exchange energies contain, in addition to the second-order
gradients of the local density and inverse effective mass, new terms
that account for the momentum dependence of the effective mass.

We have checked our Extended Thomas-Fermi approach to the Hartree-Fock
energy by performing restricted variational calculations with the
Gogny and Brink-Boeker effective forces of the binding energy of
$^{4}He$, $^{16}O$ and $^{40}Ca$ using trial harmonic oscillator
local densities. At a quantal level our Extended
Thomas-Fermi approach up to $\hbar^2$-order to the exchange energy can
be used within the Kohn-Sham scheme to obtain a local approximation to
the non-local quantal energy density. It is found that this Kohn-Sham
approach gives binding energies and root mean square radii
similar to the corresponding results obtained using the
Negele-Vautherin and Campi-Bouyssy truncations of the quantal
density matrix.

We have also performed restricted variational Strutinsky averaged
Hartree-Fock calculations with the same finite-range nuclear
interactions. These results are compared with the pure Thomas-Fermi
(Slater) and Extended Thomas-Fermi values obtained starting from the
Strutinsky kinetic energy and particle densities. The Extended
Thomas-Fermi binding energies including only $\hbar^2$-order
contributions
are not able to reproduce the Strutinsky results and consequently
cannot be used for obtaining the shell energies. We have approximately
estimated the $\hbar^4$-order contribution to the binding energy
and verified that if this correction is taken into account, the
Strutinsky values are practically recovered.

Although the Extended Thomas-Fermi approach applied to a
non-local one-body Hamiltonian gives reasonably good results, to
improve this semiclassical approximation by adding the full
$\hbar^4$-order
contributions seems to be in order. Another way of improving the
semiclassical results presented here to obtain the smooth part of the
energy is by using the Variational
Wigner-Kirkwood approach \cite{Cen} which slightly differs
from the Extended Thomas-Fermi approximation presented in
this paper. We reserve these extensions of the semiclassical
calculations in the non-local case for a future work.

On the other hand, other effective finite-range forces such as M3Y
together with the Double
Folded Model have been recently applied to compute the real part of
the microscopic heavy-ion optical potential. In these calculations,
the exchange part is usually obtained using the Negele-Vautherin or
Campi-Bouyssy approaches to the density matrix with a
semiclassical kinetic energy density \cite{Ism}. To use the full
Extended
Thomas-Fermi density matrix to obtain the real part of the heavy-ion
optical potential is another promising application of the method
developed
in this paper and will be presented in a forthcoming publication
\cite{Soub1}.

\section{Acknowledgements}

The authors are indebted to P.Schuck and M.Centelles for very useful
discussions
and with L.Egido for supplying us the full quantal results. This work
has been partially supported by Grants PB95-1249 from the DGICIT
(Spain) and 1998SGR-00011 from DGR (Catalonia).

\pagebreak
\begin{center}
{\bf Appendix}
\end{center}
The WK density matrix up to $\hbar^2$ order (assuming degeneracy g)
in coordinate space for a non-local potential
is given by the inverse Wigner transform of (\ref{eq24})
\bea
\tilde{\rho}_{WK}(\mbold{R},s) &=& g \int \frac{d \mbold{k}}{(2
\pi)^3}
\tilde{f}_{WK}(\mbold{R},k) e^{i \mbold{k} \mbold{s}} = \frac{g k_F^3}
{6 \pi^2} \frac{3 j_1 (k_F s)}{k_F s} \nonumber \\
&-& \frac{g \hbar^2}{16 m \pi^2} \frac{\partial}{\partial \lambda}
\int dk k^2 j_0 (k s) F_1 (\mbold{R},k) \delta (\lambda - H_W)
\nonumber \\
&+& \frac{g \hbar^2}{48 m \pi^2} \frac{\partial^2}{\partial \lambda^2}
\int dk k^2 j_0 (k s) F_2 (\mbold{R},k) \delta (\lambda - H_W)
\label{eqA.1} \eea
that written in terms of the local Fermi momentum
$k_F$ with the help of:
\be
\delta (\lambda - H_W) = \frac{m}{\hbar^2} \frac{\delta(k - k_F)}
{k_F f(\mbold{R},k_F)}
\label{eqA.2} \ee
and
\be
\frac{\partial \lambda}{\partial k_F} = \frac{\hbar^2 k_F}{m} +
V_k(\mbold{R},k_F) = \frac{\hbar^2 k_F f(\mbold{R},k_F)}{m}
\label{eqA.3} \ee
reads
\bea
&& \tilde{\rho}_{WK}(\mbold{R},s)  = \frac{g k_F^3}{6 \pi^2} \frac{3
j_1 (k_F s)}{k_F s}
+ j_0 (k_F s) \rho_{2,WK} (\mbold{R}) \nonumber \\ &+&
\frac{g}{48 \pi^2} \frac{\partial j_0 (k_F s)}{\partial k_F}
\frac{m}{\hbar^2 k_F f(\mbold{R},k_F)} \bigg[ \frac{m}{\hbar^2 k
f(\mbold{R},k)} \frac{d}{d k} \big[ \frac{k F_2}{f(\mbold{R},k)} \big]
+ \frac{m}{\hbar^2} \frac{d}{d k} \big[ \frac{F_2}{f(\mbold{R},k)^2}
\big] \nonumber
\\ &-& \frac{3 k F_1}{f(\mbold{R},k)} \bigg]_{k=k_F}
+ \frac{g}{48 \pi^2} \frac{\partial^2 j_0 (k_F s)}{\partial k_F^2}
\big(\frac{m}{\hbar^2 k_F f(\mbold{R},k_F)}\big)^2 \bigg[ \frac{k F_2}
{f(\mbold{R},k)} \bigg]_{k=k_F},
\label{eqA.4} \eea
where $\rho_{2,WK}$ is the WK
$\hbar^2$-order contribution to the density in the non-local problem
given in the Appendix A of \cite{Cen} and the inverse
effective mass $f(\mbold{R},k)$ is defined as in eq.(\ref{eq25}).

In eq.(\ref{eqA.4}) the gradients of the non-local potential
$V(\mbold{R},k)$ appearing in $F_1(\mbold{R},k)$,
$F_2(\mbold{R},k)$ and their momentum  derivatives have to be
evaluated at $k=k_F$.
To do this one starts from the definition of the Fermi energy:
\be
\frac{\hbar^2 k_F^2}{2 m} + V(\mbold{R},k) = \lambda,
\label{eqA4.b} \ee
where $k_F$ is also a function of $\mbold{R}$. Now taking
the gradients of (\ref{eqA4.b}), the spatial derivatives of the
potential are transformed into gradients of the local Fermi momentum
through:
\be
(\bnabla{V})_{k_F} + \frac{\hbar^2 k_F}{m} f(\mbold{R},k_F)
\bnabla{k_F} = 0
\label{eqA.5} \ee
\bea
(\Delta V)_{k_F} &+& \frac{\hbar^2}{m} \big[ k_F f(\mbold{R},k_F)
\Delta k_F + f(\mbold{R},k_F) (\bnabla k_F)^2
+ 2 k_F \bnabla{f(\mbold{R},k_F)} \bnabla{k_F} \nonumber \\
&+& k_F f_k(\mbold{R},k_F) (\bnabla k_F)^2 \big] = 0.
\label{eqA.6} \eea
To obtain the ETF DM one proceeds as in the local potential case.
First the WK local density up to $\hbar^2$-order is inverted to obtain
$k_F$
\be
k_F = k_0 - \frac{2 \pi^2}{g k_0^2} \rho_{2,WK} [k_0],
\label{eqA.7} \ee
where $k_0$ is the zeroth order local Fermi momentum given by
$k_0=(6 \pi^2 \rho/g)^{1/3}$ and the gradients of
$V$ that appear in $\rho_{2,WK}$ \cite{Cen} have been replaced by
gradients of $k_0$ with the help of eqs.(\ref{eqA.5} - \ref{eqA.6}).

Finally, expanding $k_F$ in the WK DM (\ref{eqA.4}) with the help of
(\ref{eqA.7}) one obtains the ETF density matrix written as
\bea
\tilde{\rho}_{ETF} &=& \frac{g k_0^3}{6 \pi^2} \frac{3 j_1 (k_0
s)}{k_0 s}
\nonumber \\ &+& \frac{g k_0 s}{144 \pi^2} \bigg[ \big( 7 + 6k_0 \frac
{f_k}{f} + 2 k_0^2 \frac{f_{kk}}{f} - k_0^2 \frac{f_k^2}{f^2} \big)
\frac{(\bnabla{k_0})^2}{k_0} + \big( 4 + 2 k_0 \frac{f_k}{f} \big)
\Delta k_0  \nonumber \\ &+& (6 - 2 k_0 \frac{f_k}{f})
\frac{\bnabla{f} \bnabla{k_0}}{f} +
4 k_0 \frac{\bnabla{f_k} \bnabla{k_0}}{f} + 2 k_0 \frac{\Delta
f}{f} - k_0 \frac{(\bnabla{f})^2}{f} \bigg] \frac{d j_0 (k_0 s)}{d(k_0 s)}
\nonumber \\ &+& \frac{g k_0^2 s^2}{144 \pi^2}
\bigg[ 2 \frac{(\bnabla{k_0})^2}{k_0} - \Delta k_0 \bigg]
\frac{d^2 j_0 (k_0 s)}{d (k_0 s)^2}.
\label{eqA.8} \eea
If now the gradients of $k_0$ are written in terms of gradients of the
density using eqs.(\ref{eq6a} -\ref{eq6b}) one obtains the ETF DM for
a non-local potential written as a functional of the local density
which is just (\ref{eq27}).

The HF energy of an uncharged and spin-saturated nucleus in
the ETF approximation can be obtained by replacing
the quantal integrand in (\ref{eq28}) by its corresponding ETF
approximation.
The ETF kinetic energy density can be derived from the ETF DM using
\cite{Ring}
\be
\tau_{ETF}(\mbold{R}) = \bigg(\frac{1}{4} \Delta_R - \Delta_s \bigg)
\tilde{\rho}_{ETF}(\mbold{R},s) \vert_{s=0}
\label{eqA.9} \ee
from where eq.(\ref{eq30}) is easily obtained.

The direct energy is obtained using the the diagonal part of the DM
that in the ETF approach reduces simply to the local density $\rho$.

The exchange potential is given by $V_{ex}(\mbold{R},s)=v(s)
\tilde{\rho}(\mbold{R},s)$ and, consequently, the ETF exchange energy
will be
\bea
\varepsilon_{ex}^{ETF}(\mbold{R},s) &=& \frac{1}{2} \int d\mbold{s}
V_{ex}^{ETF}(\mbold{R},s) \tilde{\rho}_{ETF}(\mbold{R},s) \nonumber \\
&=&\frac{1}{2} \int d\mbold{s} v(s) \bigg[
\tilde{\rho}_{ETF,0}^2(\mbold{R},s) + 2
\tilde{\rho}_{ETF,0}(\mbold{R},s) \tilde{\rho}_{ETF,2}(\mbold{R},s)
\bigg] \nonumber \\ &=& \frac{1}{2} \rho^2(\mbold{R}) \int d\mbold{s}
v(s) \frac{9 j_1^2 (k_F s)}{(k_F s)^2} + \int d\mbold{s}
V_{ex,0}^{ETF}(\mbold{R},s) \tilde{\rho}_{ETF,2}(\mbold{R},s).
\label{eqA.10} \eea

The integral over $\mbold{s}$ in the $\hbar^2$ contribution to the ETF
exchange energy can be performed analytically taking into account
the fact that in Wigner space the $\hbar^0$ ETF exchange potential
(Slater) can also be written as
\be
V_{ex,0}^{ETF}(\mbold{R},k) = \int d \mbold{s}
V_{ex,0}^{ETF}(\mbold{R},s)
e^{-i \mbold{k} \mbold{s}} = \int d \mbold{s} v(s) \tilde{\rho}_0
(\mbold{R}, s) j_0 (k s)
\label{eqA.11} \ee
if the exchange potential is spherically symmetric in $\mbold{k}$.

The $k$ derivatives calculated at $k=k_0$ are easily obtained
starting from(\ref{eqA.11}):
\be
V_{ex,0,k}^{ETF}(\mbold{R},k_0) = \int d \mbold{s} v(s)
\tilde{\rho}_0(\mbold{R},s) s \frac{d j_0 (k s)}{d(k s)} =
\frac{\hbar^2 k (f - 1)}{m} \vert_{k=k_0}
\label{eqA.12} \ee
\be
V_{ex,0,kk}^{ETF} (\mbold{R},k_0) = \int d \mbold{s} v(s)
\tilde{\rho}_0(\mbold{R},s) s^2 \frac{d^2 j_0 (k s)}{d(k s)^2} =
\frac{\hbar^2 (f + k f_k - 1)}{m} \vert_{k=k_0}.
\label{eqA.13} \ee
With the help of eqs.(\ref{eqA.12}) and (\ref{eqA.13}) the
integral over $\mbold{s}$ of the $\hbar^2$ part of (\ref{eqA.10}) can
be done obtaining
\bea
\varepsilon_{ex,2}(\mbold{R}) &=& \frac{\hbar^2 k_0^2}{36 m
\pi^2} \bigg\{(f - 1) \bigg[ \big( 9 + 6k_0 \frac
{f_k}{f} + 2 k_0^2 \frac{f_{kk}}{f} - k_0^2 \frac{f_k^2}{f^2} \big)
\frac{(\bnabla{k_0})^2}{k_0}
+ \big( 3 + 2k_0 \frac{f_k}{f} \big) \Delta k_0
\nonumber \\ &+& (6 - 2 k_0 \frac{f_k}{f})
\frac{\bnabla{f} \bnabla{k_0}}{f} +
4 k_0 \frac{\bnabla{f_k} \bnabla{k_0}}{f} + 2 k_0 \frac{\Delta
f}{f} - k_0 \frac{(\bnabla{f})^2}{f} \bigg] \nonumber \\ &+& k_0 f_k
\bigg[ 2 \frac{(\bnabla{k_0})^2}{k_0} - \Delta k_0 \bigg] \bigg\}.
\label{eqA.17} \eea
Usung this equation and replacing the gradients of $k_0$ by
gradients of $\rho$ with the help of (\ref{eq6a}-\ref{eq6b}) together
with the ETF kinetic energy density given by (\ref{eq30}),
the $\hbar^2$ contribution to the exchange energy in the ETF
approximation can be recast in the form given in eq.(\ref{eq33}).

\pagebreak

\pagebreak
\begin{center}
FIGURE CAPTIONS
\end{center}

Figure 1. Ratio of off-diagonal to diagonal quantal density matrix
$\rho(\mbold{R}+\mbold{s}/2,\mbold{R}-\mbold{s}/2)/\rho(\mbold{R})$
for $^{40}Ca$ as
a function of the interparticle distance $s$ for some values of the
centre-of-masss distance $R$ (in fm). The different curves
appearing in the Figure are explained in the text.

Figure 2. Ratio of off-diagonal to diagonal semiclassical density
matrix
$\rho(\mbold{R}+\mbold{s}/2,\mbold{R}-\mbold{s}/2)/\rho(\mbold{R})$
for $^{40}Ca$ as
a function of the interparticle distance $s$ for some values of the
centre-of-mass distance $R$ (in fm). The different curves
appearing in the Figure are explained in the text.

\pagebreak
\begin{center}
TABLE CAPTIONS
\end{center}

Table1. Extended Thomas-Fermi (ETF) Coulomb exchange energies for
$^{4}He$, $^{16}O$ and $^{40}Ca$ compared with the quantal (QM),
Slater (SL), Negele-Vautherin (NV) and Campi-Bouyssy (CB) values
reported in ref. \cite{CB}.

Table 2. Total quantal binding energies and root mean square radius
for $^{4}He$, $^{16}O$ and $^{40}Ca$ obtained quantally (QM) and using
the different approximations described in the text.

Table 3. Total binding energies and root mean square radius
for $^{4}He$, $^{16}O$ and $^{40}Ca$ obtained from the Strutinsky
averaged
method (ST) and with the different Extended Thomas-Fermi approaches
described in the text.
\pagebreak
\begin{center}
   TABLE I
\end{center}

\begin{center}  \small
\begin{tabular}{l c c c c c c}
\hline
  & $^{4}He$ && $^{16}O$ && $^{40}Ca$ & \\
\hline
QM
  & -0.86 && -2.98 && -7.46 \\
NV
  & -0.47 && -2.31 && -6.42 \\
CB
  & -0.78 && -2.75 && -7.03 \\
SL
  & -0.74 && -2.75 && -7.05 \\
ETF
  & -0.82 && -2.89 && -7.31 \\
\hline
\end{tabular}
\end{center}

\vspace*{1.0cm}
\begin{center}
   TABLE II
\end{center}

\begin{center}  \small
\begin{tabular}{l c c c c c c}
\hline
  & $^{4}He$ && $^{16}O$ && $^{40}Ca$ & \\
\hline
  & $E$ & $<r^2>^{1/2}$
  & $E$ & $<r^2>^{1/2}$
  & $E$ & $<r^2>^{1/2}$  \\
  & (MeV) & (fm) & (MeV) & (fm) & (MeV) & (fm) \\
\hline
QM
  & -28.24 & 1.72 & -106.62 & 2.65 & -323.98 & 3.36 \\
NV
  & -23.69 & 1.86 & -100.32 & 2.71 & -313.49 & 3.40 \\
CB
  & -25.61 & 1.80 & -101.96 & 2.70 & -315.22 & 3.39 \\
NV(KS)
  & -21.75 & 1.89 &  -97.55 & 2.73 & -309.42 & 3.41 \\
CB(KS)
  & -22.39 & 1.87 &  -98.27 & 2.72 & -310.10 & 3.40 \\
SL(KS)
  & -17.51 & 2.00 &  -87.85 & 2.79 & -291.78 & 3.45 \\
ETF(KS)
  & -23.78 & 1.82 & -101.61 & 2.69 & -317.11 & 3.38 \\
\hline
  & $^{4}He$ && $^{16}O$ && $^{40}Ca$ & \\
\hline
  & $E$ & $<r^2>^{1/2}$
  & $E$ & $<r^2>^{1/2}$
  & $E$ & $<r^2>^{1/2}$  \\
  & (MeV) & (fm) & (MeV) & (fm) & (MeV) & (fm) \\
\hline
QM
  & -29.66 & 1.88 & -138.07 & 2.63 & -406.47 & 3.34 \\
NV
  & -28.53 & 1.91 & -135.72 & 2.65 & -402.68 & 3.36 \\
CB
  & -29.04 & 1.90 & -136.31 & 2.64 & -403.22 & 3.36 \\
NV(KS)
  & -27.41 & 1.93 & -133.65 & 2.65 & -399.57 & 3.36 \\
CB(KS)
  & -27.53 & 1.93 & -133.80 & 2.65 & -399.62 & 3.36 \\
SL(KS)
  & -25.80 & 1.96 & -129.49 & 2.67 & -392.33 & 3.37 \\
ETF(KS)
  & -29.05 & 1.91 & -135.38 & 2.64 & -402.66 & 3.35 \\
\hline
\end{tabular}
\end{center}
\pagebreak
\begin{center}
   TABLE III
\end{center}

\begin{center}  \small
\begin{tabular}{l c c c c c c}
\hline
  & $^{4}He$ && $^{16}O$ && $^{40}Ca$ & \\
\hline
  & $E$ & $<r^2>^{1/2}$
  & $E$ & $<r^2>^{1/2}$
  & $E$ & $<r^2>^{1/2}$  \\
  & (MeV) & (fm) & (MeV) & (fm) & (MeV) & (fm) \\
\hline
ST
  & -16.71 & 1.94 &  -89.15 & 2.75 & -296.79 & 3.42 \\
SL
  & -22.77 & 1.94 &  -97.73 & 2.75 & -307.94 & 3.42 \\
ETF(a)
  & -22.51 & 1.94 &  -98.06 & 2.75 & -310.51 & 3.42 \\
ETF(b)
  & -21.71 & 1.94 &  -96.20 & 2.75 & -307.03 & 3.42 \\
ETF($\hbar^4$)
  & -17.64 & 1.94 &  -90.19 & 2.75 & -298.71 & 3.42 \\
\hline
  & $^{4}He$ && $^{16}O$ && $^{40}Ca$ & \\
\hline
  & $E$ & $<r^2>^{1/2}$
  & $E$ & $<r^2>^{1/2}$
  & $E$ & $<r^2>^{1/2}$  \\
  & (MeV) & (fm) & (MeV) & (fm) & (MeV) & (fm) \\
\hline
ST
  & -22.95 & 1.95 & -125.14 & 2.67 & -386.44 & 3.37 \\
SL
  & -31.85 & 1.95 & -140.59 & 2.67 & -409.45 & 3.37 \\
ETF(a)
  & -27.78 & 1.95 & -133.39 & 2.67 & -398.18 & 3.37 \\
ETF(b)
  & -27.60 & 1.95 & -132.85 & 2.67 & -397.22 & 3.37 \\
ETF($\hbar^4$)
  & -23.67 & 1.95 & -125.30 & 2.67 & -386.07 & 3.37 \\
\hline
\end{tabular}
\end{center}

%
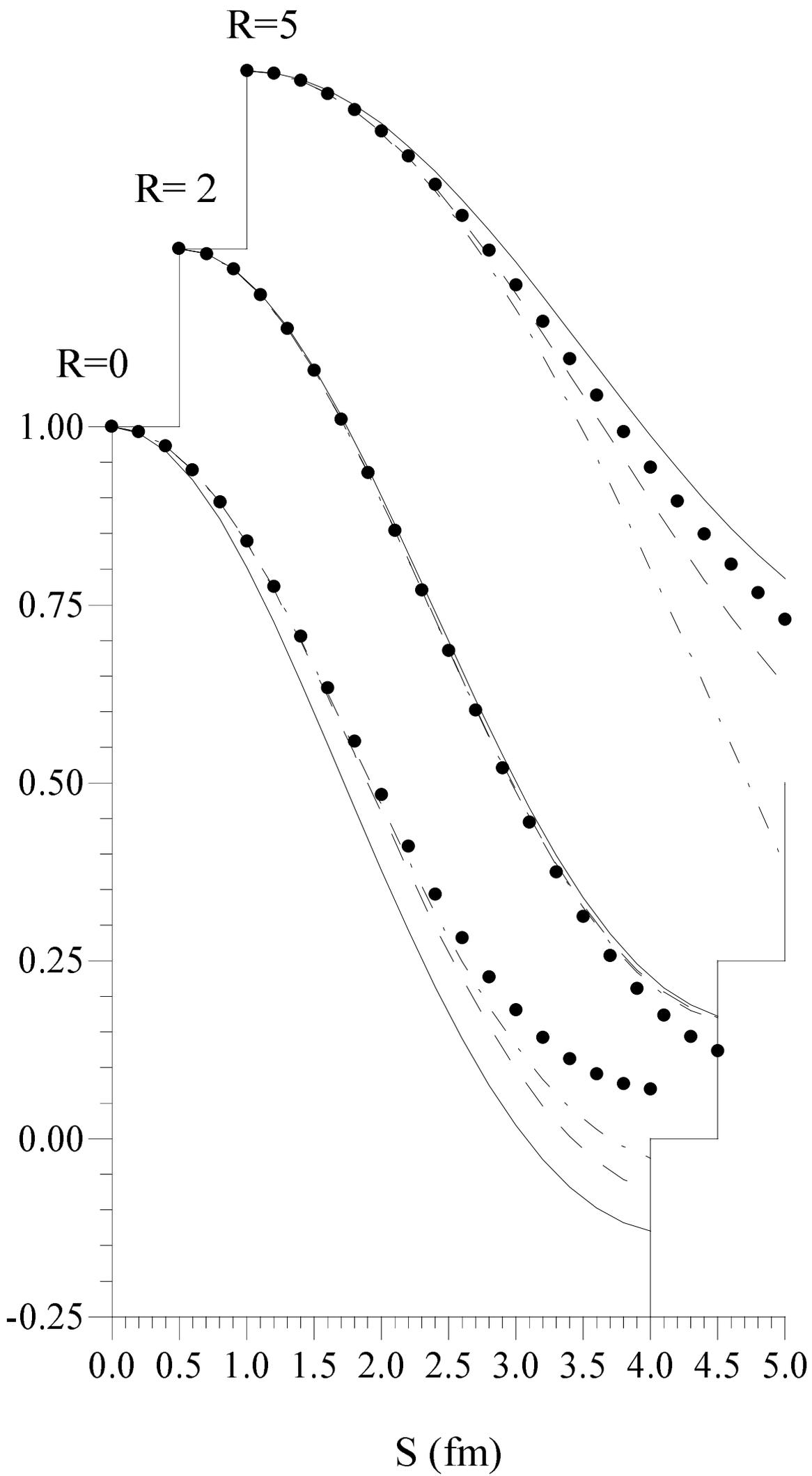
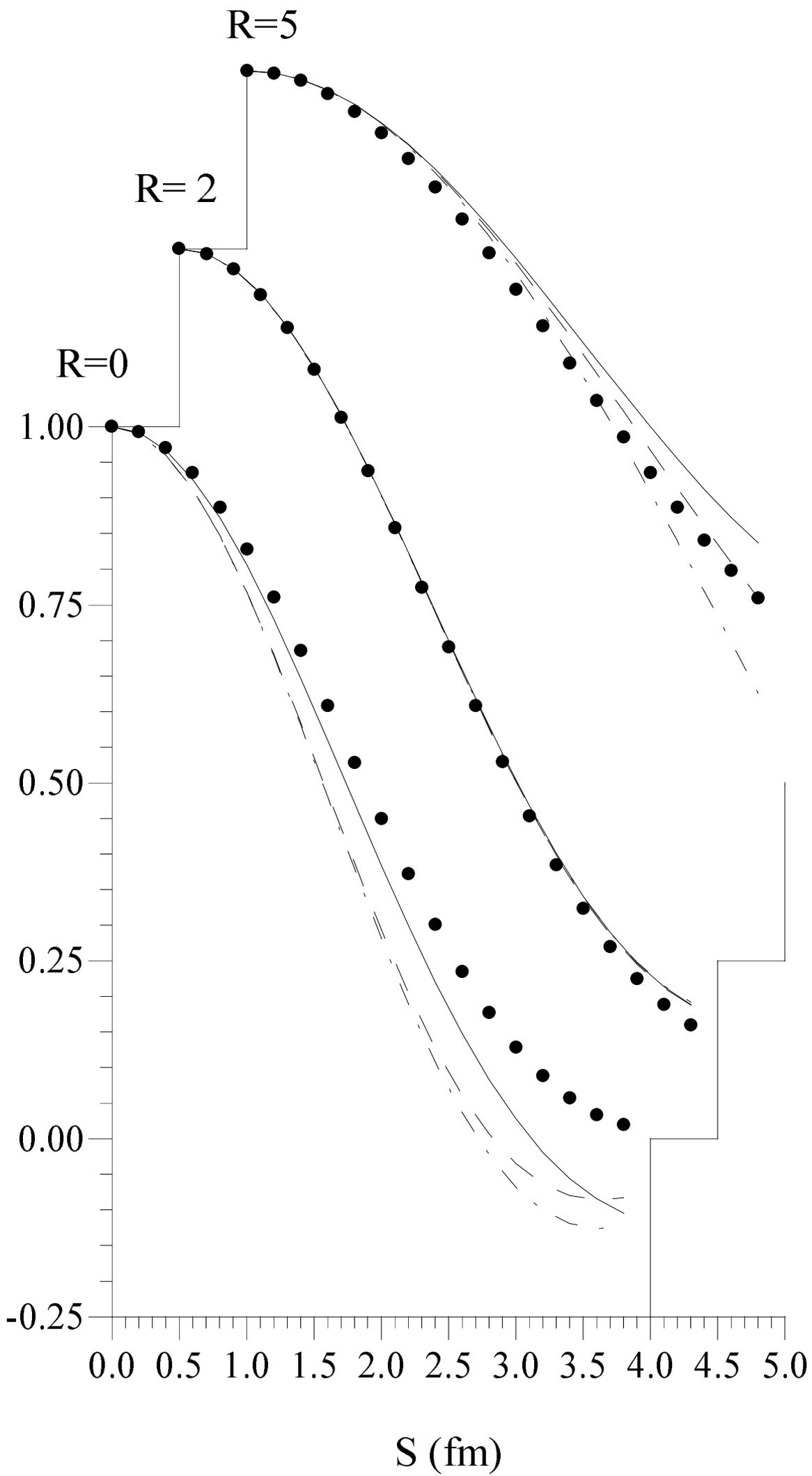
\end{document}